\documentclass[11pt]{article}
\usepackage{amsmath,amssymb,color,graphics,epsfig,cite}

\usepackage{amsfonts}

\newcommand{\be}{\begin{equation}}
\newcommand{\ee}{\end{equation}}
\newcommand{\bea}{\setlength\arraycolsep{2pt} \begin{eqnarray}}
\newcommand{\eea}{\end{eqnarray}}
\newcommand{\nn}{\nonumber}

\def\ft#1#2{{\textstyle{\frac{\scriptstyle #1}{\scriptstyle #2} } }}
\def\fft#1#2{{\frac{#1}{#2}}}

\def\0{{\sst{(0)}}}
\def\1{{\sst{(1)}}}
\def\2{{\sst{(2)}}}
\def\3{{\sst{(3)}}}
\def\4{{\sst{(4)}}}
\def\5{{\sst{(5)}}}
\def\6{{\sst{(6)}}}
\def\7{{\sst{(7)}}}
\def\8{{\sst{(8)}}}
\def\sst#1{{\scriptscriptstyle #1}}

\begin{document}


\begin{center}
{\Large {\bf Holographic Complexity and Two Identities of Action Growth}}

\vspace{40pt}
{\bf Hyat Huang, Xing-Hui Feng and  H. L\"u}

\vspace{10pt}

{\it Center for Advanced Quantum Studies, Department of Physics, \\
Beijing Normal University, Beijing 100875, China}

\vspace{40pt}

\underline{ABSTRACT}
\end{center}

The recently proposed complexity-action conjecture allows one to calculate how fast one can produce a quantum state from a reference state in terms of the on-shell action of the dual AdS black hole at the Wheeler-DeWitt patch.  We show that the action growth rate is given by the difference of the generalized enthalpy between the two corresponding horizons. The proof relies on the second identity that the surface-term contribution on a horizon is given by the product of the associated temperature and entropy.

\vfill {\footnotesize hyat@mail.bnu.edu.cn\ \ \ xhfengp@mail.bnu.edu.cn \ \ \ mrhonglu@gmail.com}

\thispagestyle{empty}

\pagebreak




\section{Introduction}

Holographic principle \cite{'tHooft:1993gx,Susskind:1994vu}, the AdS/CFT correspondence \cite{Maldacena:1997re} in particular, provides a powerful tool to study a strongly-coupled quantum theory at the boundary using a highly classical theory in the bulk.  One area of research with widespread interest is relating \cite{Harlow:2013tf,Susskind:2013aaa,Susskind:2014rva,Stanford:2014jda} the quantum computational complexity \cite{watrous}, the minimum number of elementary operations needed to produce a state of interest from a reference state, to black hole physics. The most recent proposal is the complexity-action (CA) conjecture that the quantum complexity ${\cal C}$ of a boundary state is related to the corresponding bulk action ${\cal A}$ in the region called the Wheeler-DeWitt patch \cite{Brown:2015bva,Brown:2015lvg}, namely
\be
{\cal C}=\fft{\cal A}{\pi\hbar }\,.
\ee
This implies that how fast information can be stored may be computed by the growth rate of the on-shell action of the corresponding black hole.

The action of the Wheeler-DeWitt patch for anti-de Sitter (AdS) black holes is essentially evaluated over the spacetime volume between the outer and inner horizons \cite{Brown:2015lvg}. (See \cite{Lehner:2016vdi,Chapman:2016hwi} for further discussion on the global structure of the Wheeler-DeWitt patch.) The action growth for stationary AdS black holes with various charge or rotation parameters was computed \cite{Brown:2015lvg,Cai:2016xho}.  For a variety of single-charged and/or single-rotation black holes, the answer takes the form \cite{Cai:2016xho}
\be
\fft{d{\cal A}}{dt} = (M - \Omega J - \mu Q)_+ - (M-\Omega J -\mu \Omega)_-\,.\label{spcase1}
\ee
We have checked a great many further examples of AdS black holes in literature, including the static and rotating black holes in gauged STU models, and Kerr-AdS black holes with multiple rotations in general dimensions
\cite{Behrndt:1998jd,Duff:1999gh,Cvetic:1999xp,Hawking:1998kw,Gibbons:2004js,%
Gibbons:2004uw,Chong:2005hr,Wu:2011gq,Lu:2013eoa}. The general formula takes the form
\be
\fft{d{\cal A}}{dt} = (M - \Omega^i J^i - \mu^\alpha Q^\alpha)_+ - (M-\Omega^i J^i -\mu^\alpha \Omega^\alpha)_-\,,
\label{spcase2}
\ee
where the repeated indices imply summation.  The large number of examples we have checked indicate the formula is robust.  The motivation of this paper is to give a formal proof.
To do so, we find that the cumbersome formula (\ref{spcase2}) can be further abstracted to be
\be
\fft{d{\cal A}}{dt} = (F+TS)_+ - (F+TS)_-= {\cal H}_+ - {\cal H}_-\,,\label{generalcase}
\ee
where $F$ is the free energy obtained from the Euclidean action via the quantum statistic relation (QSR) \cite{Gibbons:1976ue}, and ${\cal H} \equiv F + TS$ is the generalized enthalpy, whose terminology will be justified later.

Assuming that the QSR holds, the key to prove (\ref{generalcase}) is then the identity that the surface contribution to the action growth at each horizon is precisely the product of the associated Hawking temperature and entropy, namely
\be
\fft{d{\cal A}^{\rm surf}}{dt}\Big|_\pm = T_\pm  S_\pm\,.\label{id2}
\ee

The paper is organized as follows.  In section 2, we establish the identities in two-derivative Einstein gravities.  In section 3, we establish them in general higher-derivative gravities.  We conclude the paper and give further discussions in section 4.

\section{Action growth in Einstein gravity}

We begin with Einstein gravity with minimally-coupled matter in general $D=n+1$ dimensions.  The action can be expressed as $\int dt\, L$, where the Lagrangian $L$ consists of the bulk and boundary terms.  We are interested in stationary black holes for which the on-shell Lagrangian $L$ is time-independent.  In other words, we have $\fft{d{\cal A}}{dt}=L$, with
\bea
L^{\rm bulk}&=&\fft{1}{16\pi}\int_{\cal M} d^n x\, {\cal L}=\fft{1}{16\pi}\int_{\cal M} d^n x\, \Big(\sqrt{-g} R - {\cal L}^{\rm mat}\Big)\,,\nn\\
L^{\rm surf}&=&L^{\rm GH} + L^{\rm ct}\,,\qquad L^{\rm GH}=\fft{1}{8\pi} \int_{\partial {\cal  M}} d^{n-1}x\, \sqrt{-h} K\,.
\eea
Here $K=h^{\mu\nu} K_{\mu\nu}$ is the trace of the second fundamental form $K_{\mu\nu}=h_\mu{}^\rho \nabla_\rho n_\nu$ and $h_{\mu\nu}=g_{\mu\nu}-n_\mu n_\nu$, with $n^\mu$ being the unit vector normal to the surface \cite{Gibbons:1976ue}. (Note that the cosmological constant $\Lambda$ belongs to ${\cal L}^{\rm mat}$ in this paper.) For asymptotically AdS backgrounds, it is also necessary to introduce the counter terms \cite{Emparan:1999pm}
\bea
L^{\rm ct} &=& \fft{1}{16\pi} \int_{\partial {\cal M}} d^{n-1} \sqrt{-h}\Big[
-\fft{2(n-3)}{\ell} + \fft{\ell}{(n-4)} {\cal R}\cr
&&\qquad\qquad + \fft{\ell^3}{(n-6)(n-4)^2}({\cal R}^{\mu\nu} {\cal R}_{\mu\nu} -
\fft{n-2}{4(n-3)} {\cal R}^2) + \cdots\Big]\,.\label{Lct}
\eea
where ${\cal R}^{\mu\nu\rho\sigma}$ and its contraction denote curvatures in the boundary metric $h_{\mu\nu}$, and $\ell$ is the AdS radius.

The QSR states that for black holes, the on-shell Euclidean action is $I_{\rm E}=F/T$, where the temperature $T$ is the inverse of the period of the Euclidean time, and $F$ is the thermodynamical free energy of the black holes \cite{Gibbons:1976ue}. To be specific, the QSR implies
\be
-F= \fft{1}{16\pi}\int_+^\infty d^n x {\cal L} + L^{\rm GH}_\infty + L^{\rm ct}_\infty\,.\label{fe}
\ee
For Euclideanized black holes, there is only one boundary, located at the asymptotic infinity.  The Euclidean Killing horizon is not a boundary but the middle of the bulk.

    In order to compute the action growth of a black hole, we need to evaluate it in the original Minkowski signature.  Since the event horizon is not geodesically complete, we need to count also the boundary contribution on the horizon.  It is clear that all the polynomial invariants of ${\cal R}^{\mu\nu\rho\sigma}$ in (\ref{Lct}) are finite and hence $L^{\rm ct}$ vanishes since $\sqrt{-h}$ vanishes on the horizon.  Thus the on-shell action on and out of the horizon is
\be
L_+ = \fft{1}{16\pi}\int_+^\infty d^n x\, {\cal L} + L^{\rm surf}_\infty - L^{\rm surf}_+
=-F - L^{\rm GH}_+\,.\label{Lplus}
\ee
The most general near-horizon geometry up to the relevant order takes the form
\bea
ds^2 &=& V \Big(\fft{dr^2}{4\pi T (r-r_0)} - 4\pi T (r-r_0) dt^2\Big) + g_{ij} (dy^i-\omega^i dt) (dy^j - \omega^j dt)\,,\nn\\
V&=&V(y) + {\cal O}(r-r_0)\,,\quad g_{ij} = g_{ij}^0(y) + {\cal O}(r-r_0)\,,\quad
\omega^i=(\omega^0)^i + {\cal O}(r-r_0)\,.\label{nearh}
\eea
It is then straightforward to evaluate that
\be
L^{\rm GH}_+ = T S\,,\qquad S=\ft14 \int  d^{n-1} y\, \sqrt{\det(g_{ij}^0)}\,.\label{LGHplus}
\ee
Here $S$, one-quarter of the horizon area, is precisely the Bekenstein-Hawking entropy.  It is rather subtle to evaluate the boundary terms on the null surfaces like horizons approaching from the inside, and new contributions on the null surfaces were introduced in \cite{Lehner:2016vdi,Chapman:2016hwi}.  Analogous results of (\ref{LGHplus}) involving the new contributions were also obtained in \cite{Couch:2016exn,Yang:2016awy}. We shall comment on our approach presently.

For black holes that have the ``inner'' as well as the usual ``outer'' horizons, the first law of black hole ``thermodynamics'' is formally valid for both horizons.  For example, the Kerr-Newman-(AdS) black hole has two horizons, and we may label the quantities associated outer and inner horizons with ``$+$'' and ``$-$'' subscripts. The first law at each horizon takes the same form
\be
dM = T_\pm dS_\pm + \mu_\pm dQ + \Omega_\pm dJ + V_\pm dP\,.
\ee
Here the pressure $P=-(D-2)\Lambda/(16\pi)$. For this reason, $M$ is more appropriately called the enthalpy instead of the energy of AdS black holes \cite{Kastor:2009wy,Cvetic:2010jb}. The free energy associated with each horizon for the Kerr-Newman-(AdS) black hole can be formally computed using the QSR, giving rise to the Lagrangian in Minkowski signature as
\be
L_\pm = \fft{1}{16\pi}\int_\pm^\infty d^n x\, {\cal L} + L^{\rm surf}_\infty - L^{\rm surf}_\pm =-F_\pm - L^{\rm GH}_\pm = -(M - \mu_\pm Q - \Omega_\pm J)\,.
\ee
In general the formulae (\ref{Lplus}) and (\ref{LGHplus}) associated with the outer horizon can be generalized to be valid for both horizons, yielding
\be
L_\pm = -F_\pm - T_\pm S_\pm=-{\cal H}_\pm \,.
\ee
We refer to ${\cal H}$ as generalized enthalpy, since it is related to the enthalpy $M$ by some Legendre transformation that does not involve either $(T,S)$ or $(P,V)$. It follows that the Lagrangian of the Wheeler-DeWitt patch is given by
\be
L^{\rm WD}= L_- - L_+ = \fft{1}{16\pi} \int_-^+ d^n x\, {\cal L} + L^{\rm GH}_+ -  L^{\rm GH}_- = {\cal H}_+ -{\cal H}_-\,.\label{Lpm}
\ee
We thus prove the identity (\ref{generalcase}). We now comment on our approach of evaluating the Gibbons-Hawking surface term.  It follows from (\ref{nearh}) that in the inner and outer horizons, we have $T_-<0$ and $T_+>0$ respectively,\footnote{One may also adopt the convention that $T$ is always chosen to be positive by modifying the first law, namely $dM=\pm T_\pm dS+ \cdots$.} we choose to approach the horizon surfaces by taking the limit $r-r_\pm\rightarrow \pm 0$ respectively, in which cases the Gibbons-Hawking term is always evaluated on the time-like surfaces.  It is clear that this is a smooth limit for both the bulk action and the Gibbons-Hawking term.

The conclusion holds also for theories with non-minimally coupled matter.  As a concrete example, we consider the Brans-Dick theory:
\be
L^{\rm bulk}=\fft{1}{16\pi}\int d^{n} x\, \sqrt{-g}\, \phi R + \cdots\,,\qquad
L^{\rm surf} = \fft{1}{8\pi}\int d^{n-1}x\, \sqrt{-h}\, \phi K\,.
\ee
It is then straightforward to see that on the horizon with the near-horizon geometry (\ref{nearh}) we have
\be
L^{\rm surf}_{r=r_0} = T \times \fft{\phi(r_0)}4 \int d^{n-1} y \sqrt{\det(g_{ij}^0)}\,.
\ee
This is precisely the product of the temperature and entropy, which then leads directly to statement (\ref{generalcase}).

\section{Higher derivative gravities}

We now consider general classes of covariant gravities that are constructed from polynomial invariants of Riemann and matter tensors.  Assuming that the QSR holds for black holes in these theories, it follows from the previous discussion that the key to establish (\ref{generalcase}) is the identity (\ref{id2}).   The proof of (\ref{id2}) may appear to be difficult since the entropy in the general theory is expected to
be given by the Wald entropy formula \cite{wald}
\be
S=-\fft{1}{8} \int_{\rm horizon} d^{n-1}x\sqrt{\hat h}\, \epsilon^{ab}\epsilon^{cd} \fft{\partial \hat L}{\partial R^{abcd}}\,,\label{waldentropy}
\ee
where $\hat L$ is defined by the bulk Lagrangian as $L^{\rm bulk}=\int d^nx\, \sqrt{-g}\, \hat L$, and $\epsilon^{ab}$ is the binormal to the bifurcation surface.  For spherically-symmetric black holes in Einstein-Gauss-Bonnet theory, we find that the identity (\ref{id2}) can be shown using the results presented in \cite{Cai:2016xho}.  For general theories with minimally-coupled matter, the surface term was obtained \cite{Deruelle:2009zk} , given by
\be
L^{\rm surf} = \fft1{8\pi} \int_{\cal\partial M} d^{n-1}x\, \sqrt{-h} \fft{\partial \hat L}{\partial R^{abcd}}\, K^{ac}\, n^b n^d\,.
\ee
We expect this formula may also hold for theories with non-derivative matter couplings to curvatures, since then the matter fields can be treated as constants in the relevant terms.
Substituting the near-horizon geometry (\ref{nearh}) into the above, it is clear that approaching from the outside of the horizon, we have $L^{\rm surf}_{r\rightarrow r_0}=T S$ where $S$ is given by (\ref{waldentropy}).  The identity (\ref{generalcase}) then follows directly. (It is interesting to note that integrating over Euclidean time of $L^{\rm surf}_+$ gives precisely the entropy, providing a new method of computing the entropy.)

\section{Conclusions and discussions}

In this paper, we showed for general covariant theories that the bulk Lagrangian for stationary black holes within the inner and outer horizons satisfied
\be
(L^{\rm bulk})^+_- \equiv \int^+_- d^n x\, {\cal L}^{\rm bulk} = F_+ - F_-\,.
\ee
Furthermore, we found that the surface contribution on each horizon took the form
\be
L^{\rm surf}_\pm = T_\pm S_\pm\,.
\ee
(The identities are valid for both asymptotically AdS or flat black holes.)
Together, they give rise to the growth rate of the action of Wheeler-DeWitt patch, given by (\ref{generalcase}).  The validity of the identities relies on the two assumptions.  The first is that the QSR is valid and the second is that the Wald entropy formula correctly computes the entropy of the black hole.  Both assumptions could become problematic in higher-derivative theories with non-minimally coupled derivative matters, such as Horndeski gravity \cite{Feng:2015oea,Feng:2015wvb}.  It is of great interest to investigate these theories in this context.

One test of the CA conjecture is to compare the bound for information storage in computational science, with that of black holes, since black holes are expected to be the fastest computers \cite{Brown:2015bva}.  By studying the thermofield double states, the bound was proposed in \cite{Brown:2015bva}; it can be paraphrased for the general case as
\be
\fft{d{\cal A}}{dt} = (F+TS)_+ - (F+TS)_- \le 2(F+TS)_+ - 2 (F+TS)^{\rm gs}\,,\label{bound}
\ee
where the superscript ``gs'' denotes some appropriate ground state.  In the limit of the neutral and static black holes with only single horizon, the ground state is the AdS vacuum.  The bound is indeed saturated by the Schwarzschild-AdS black holes.  For charged or rotating black holes with two horizons, the ground state is naturally the (zero-temperature) extremal black hole of the same charge and/or angular momenta. For example, the extremal Reissner-Nordstr\o m-AdS black hole with horizon radius $r_0$ has ${\cal H}^{\rm ext}=(F+TS)^{\rm ext}=-r_0^3/\ell^2$.  The bound (\ref{bound}) however can be violated for small black holes ($r_0\ll\ell$). One explanation given in  \cite{Brown:2015bva} is that stringy effects may not be ignored for small black holes.  We find that in the asymptotically-flat limit ($\ell\rightarrow \infty$), $H^{\rm ext}=0$ and the bound (\ref{bound}) is saturated precisely.  In this limit, the black hole mass $M$ can be correctly interpreted as thermal internal energy.  For AdS black holes, on the other hand, $M$ should be interpreted as enthalpy, since the cosmological constant acts as the pressure of the system.  The relation between complexity and black hole volume was discussed in \cite{Couch:2016exn,Chapman:2016hwi}. These leads to a tantalizing possibility that the volume and pressure of the AdS black holes may play a role in resolving puzzle of violation of the complexity bound.

\section*{Acknowlegement}

The work is supported in part by NSFC grants NO. 11175269, NO. 11475024 and NO. 11235003.

\end{document}